\documentclass[nofootinbib, aps, prd, reprint, superscriptaddress]{revtex4-2}
\pdfoutput=1

\usepackage{amsmath}
\usepackage{tensor}
\usepackage{graphicx}
\usepackage{array}
\usepackage[normalem]{ulem}
\usepackage{cancel}
\usepackage{xcolor}
\usepackage{slashed}
\usepackage{afterpage}
\usepackage{appendix}
\usepackage{multirow}
\usepackage{physics}
\usepackage{float}
\usepackage{mathtools}
\mathtoolsset{centercolon}
\usepackage{multirow}
\usepackage{comment}
\usepackage{multirow}
\usepackage[dvipsnames]{xcolor}

\usepackage[colorlinks=true,linkcolor=blue,citecolor=blue,urlcolor=blue]{hyperref}

\begin{document}
\title{Naturalness and Fisher Information}

\author{James Halverson}
\email{j.halverson@northeastern.edu}
\affiliation{The NSF AI Institute for Artificial Intelligence and Fundamental Interactions\\ \smallskip}
\affiliation{Department of Physics, Northeastern University, Boston, MA 02115, USA \\ \smallskip}

\author{Thomas R.\ Harvey}
\email{trharvey@mit.edu}
\affiliation{The NSF AI Institute for Artificial Intelligence and Fundamental Interactions\\ \smallskip}
\affiliation{Center for Theoretical Physics, Massachusetts Institute of Technology, 77 Massachusetts Avenue, Cambridge, MA 02139, USA\\ \smallskip}

\author{Michael Nee }
\email{mnee@fas.harvard.edu}
\affiliation{Jefferson Physical Laboratory, Harvard University, \\ 17 Oxford Street, Cambridge, MA 02139, USA}

\begin{abstract}
Fine-tuning and naturalness, the sensitivity of low-energy observables to small changes in the fundamental parameters of a theory, are cornerstones of physics beyond the Standard Model. We propose a new measure of fine-tuning based on information theory. To each point in parameter space we associate a probability distribution over observables. Divergence measures encode the sensitivity of observables to model parameters and determine a Riemannian metric on parameter space. By Chentsov's theorem, the physically motivated metric is the Fisher information metric, up to scaling. We propose a rescaled fine-tuning matrix $\mathcal{F}_{ij}$ derived from the Fisher information matrix, whose non-zero eigenvalues serve as our measure of fine-tuning. When the number of observables exceeds the number of parameters, $\mathcal{F}_{ij}$ admits a natural geometric interpretation as the pullback of the Euclidean metric from observable space to the submanifold of admissible predictions, with large eigenvalues corresponding to highly stretched directions and indicative of fine-tuning. Our measure reproduces the familiar Barbieri--Giudice criterion as a special case, while generalising it to multiple correlated parameters. We illustrate its behaviour on dimensional transmutation, the Wilson--Fisher fixed point, a simple model of the hierarchy problem, and the electron Yukawa coupling, finding agreement with physical intuition in each case.
\end{abstract}

\maketitle


\newpage

\section{Introduction}

Naturalness corresponds to the notion that the fundamental parameters of a theory should not be fine-tuned against each other to reproduce experimental observables, and has been a guiding principle driving particle physics model-building~\cite{Susskind:1978ms}. It has led to the development of models of supersymmetry~\cite{Dimopoulos:1981zb, Nilles:1983ge, Haber:1984rc}, compositeness~\cite{Dimopoulos:1979es, Kaplan:1983sm}, cosmological selection~\cite{Weinberg:1988cp, Susskind:2003kw, Arkani-Hamed:2004ymt} and extra dimensions~\cite{Arkani-Hamed:1998jmv, Randall:1999ee} in attempts to construct natural extensions of the Standard Model (SM). Despite the importance of naturalness to particle physics, we lack a consensus definition.

There have been many attempts at constructing such a measure, originating with ref.~\cite{Ellis:1986yg, Barbieri:1987fn} which proposed using $\sqrt{c(\theta)}$, where:
\begin{align}
    c(\theta) = \left(\frac{\partial \log X(\theta)}{\partial \log \theta}\right)^2 \, ,
    \label{eq:GB-measure}
\end{align}
as a measure of fine-tuning of a parameter $\theta$ for an observable $X(\theta)$. This measure has subsequently been used to fit supersymmetric models to experimental data while including a `naturalness prior' that favours low values of $c(\theta)$~\cite{Allanach:2006jc, Cabrera:2008tj}. Alternative approaches have stressed the importance of referring to distributions on parameter space, which aligns more closely with intuitive notions of naturalness based on regions of parameter space rather than specific parameter values. 
This is achieved either by averaging $c(\theta)$ over a given region of parameter space~\cite{Anderson:1994dz, Anderson:1994tr, Athron:2007ry} or using Bayesian evidence as the basis for naturalness measures~\cite{Fichet:2012sn, Fowlie:2014xha, Fowlie:2024nhs}. 
The Bayesian measure generalises the Barbieri-Giudice measure $c(\theta)$ in certain cases~\cite{Fichet:2012sn}, and has also proved useful for carrying out phenomenological studies~\cite{Kim:2013uxa, Clarke:2016jzm}.
The Bayesian approach defines a global measure of how natural a theory is, rather than information on how fine-tuned individual parameters are.

In this work we instead construct a fine-tuning measure based on Fisher information, which will allow us to determine if individual parameters of a UV theory are fine-tuned. We will show that it reproduces intuitive notions of naturalness by applying it to models of dimensional transmutation such as QCD, the $O(N)$ model at the Wilson-Fisher fixed point~\cite{Wilson:1971dc}, the EFT of a light scalar after integrating out a heavy scalar (a simplified version of the doublet-triplet splitting problem, or the fine tuning of the Higgs mass), and fermion Yukawa couplings (technical naturalness~\cite{tHooft:1979rat}). Information-theoretic methods have previously been applied to study the geometry of theory space, including the structure of RG flow~\cite{Balasubramanian:2014bfa, Fowler:2021oje, Gaite:1996vk, Maity:2015rfa,Casini:2016udt, Cotler:2022fze, Berman:2022uov} and distances in the space of effective field theories~\cite{Stout:2021ubb, Stout:2022phm}, where relative entropy and Fisher information characterise how distributions over fields vary across theory space.

Rather than studying the geometry of theory space itself, we use information theory to quantify the sensitivity of IR observables to UV parameters, which we identify with fine-tuning. To do so, we associate to each point in parameter space $\{\theta^i : i = 1, \ldots, N\}$ a probability distribution $\rho_\theta(x)$ over dimensionless observables $\{x^a : a = 1, \ldots, n\}$. The most immediate choice of distribution would be the experimental distribution over observables. However, this carries a subtle danger: a quantity could be deemed fine-tuned simply because it has been measured to high precision, conflating experimental accuracy with theoretical naturalness. We instead consider a theoretical $\rho_\theta(x)$: if the map $X^a$ from UV parameters to IR observables is deterministic, the natural choice is the Dirac distribution $\rho_\theta(x) = \delta^{(n)}(x^a - X^a(\theta))$, sharply localised on the model's predictions\footnote{Marginalisation over UV parameters or finite-temperature effects may broaden this distribution in practice. Henceforth we allow ourselves to consider the case that the regulator is either due to a physical broadening or a by-hand regulation of the Dirac distribution.}. This is too singular to work with directly, and so we work with a finite-variance density, where the variance $\sigma$ serves as a regulator used to desingularise the $\delta$-distribution. A general fine-tuning measure must be well-behaved when the regulator is removed, since either smooth or singular $\rho_\theta(x)$ may arise physically.

Two natural choices present themselves. The first is a uniform distribution on an $\epsilon$-ball centred on the prediction,
\begin{equation}
    \rho_\theta(x) = \begin{cases} 1/\mathrm{Vol}_{S^n}(\epsilon) & \text{if } |\vec x - \vec X(\theta)| < \epsilon, \\ 0 & \text{otherwise,} \end{cases}
\end{equation}
and the second is a Gaussian,
\begin{equation}
    \rho_\theta(x) = \prod_{a=1}^n \frac{1}{\sqrt{2\pi\sigma^2}} \exp\!\left(-\frac{(x^a - X^a(\theta))^2}{2\sigma^2}\right).
\end{equation}
Both approaches are morally equivalent, differing only in how we define the near vicinity of the predicted value, and both can be viewed as different regularisations of the theoretical delta function distribution. A key requirement we shall impose is that any fine-tuning measure derived from $\rho_\theta$ must be independent of the regulator $\epsilon$ or $\sigma$. 
These parameters encode only the precision to which the model is required to fit the data, and a physically meaningful notion of naturalness should not depend on them. We shall work with the Gaussian from here on, as it will simplify our analysis.

Naturalness arises by comparing the predictions of $\rho_\theta(x)$ at nearby points $\theta$ and $\theta + \delta\theta$ via an appropriate measure of distance (or divergence) between probability distributions. If small changes in parameters produce easily distinguishable distributions, the model is highly sensitive to the precise values of its parameters, which is the hallmark of fine-tuning. Conversely, if $\rho_\theta(x)$ and $\rho_{\theta + \delta\theta}(x)$ are nearly indistinguishable, no fine-tuning is required. Expanding such a divergence to second order in $\delta\theta$ yields a metric on parameter space whose eigenvalues we use as our measure of fine-tuning. As we shall discuss in section~\ref{sec:Fisher}, this metric is largely independent of the choice of divergence, and under appropriate assumptions reduces to the Fisher Information Matrix~\cite{amari2016information}. Our measure is related to the eigenvalues of this matrix with the regulator dependence removed.

In the case where $n \geq N$, the predictions $X^a(\theta)$ trace out an $N$-dimensional submanifold of observable space, representing the locus of observations admissible by the model. We shall see that when $\rho_{\theta}$ is Gaussian our fine-tuning measure admits a natural geometric interpretation in this setting. It arises as the pullback of the Euclidean metric on observable space to this submanifold, with large eigenvalues corresponding to directions along which the submanifold is highly stretched. Furthermore, the Barbieri-Giudice measure can be recovered as a special case of our measure.

In the above discussion we have been somewhat cavalier about the choice of parameters for a model. Naturalness is not a property of a model alone, but of a model together with a prior on its parameters. This prior is usually motivated by EFT arguments, and our measure assumes that the parameters are chosen such that they are all drawn from the same prior. This point will be discussed further in section~\ref{sec:Fisher}, and returned to in the examples of section~\ref{sec:example}.

We begin by introducing the necessary theoretical background and our fine-tuning matrix $\mathcal{F}_{ij}$ in section~\ref{sec:Fisher}. In section~\ref{sec:example} we apply this measure to several examples of fine-tuning, demonstrating agreement with standard physical intuition, before concluding in section~\ref{sec:conc}.

\section{Fisher Information and Fine-Tuning}\label{sec:Fisher}

\begin{figure}[t]
    \centering
    \includegraphics[width=0.5\textwidth]{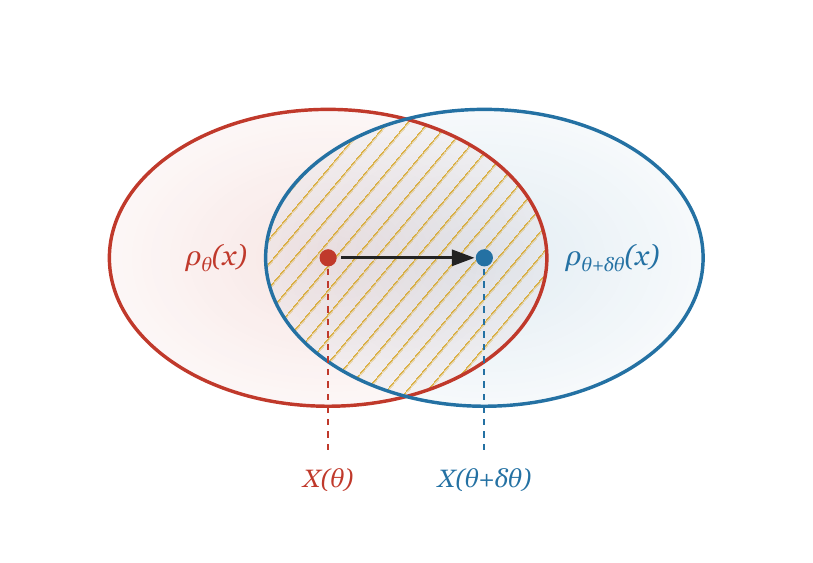}
    \caption{The distributions $\rho_\theta(x)$ and $\rho_{\theta+\delta\theta}(x)$ are tightly concentrated around the model predictions $X(\theta)$ and $X(\theta + \delta\theta)$ respectively. The degree of overlap between the two distributions, measured by the JS divergence, encodes the sensitivity of the observables to small changes in the model parameters.}
    \label{fig:distributions}
\end{figure}

\subsection{General Considerations}

Before discussing measures of fine-tuning, let us first introduce some considerations from information theory that will be relevant for our discussion, for a general $\rho_\theta(x)$. 
Classically, in information theory one wishes to quantify how much the density has changed via a change of parameters, measured by a scalar quantity $D(p_\theta || p_{\theta'})$, where the $||$ in place of a comma emphasises that $D$ need not be symmetric under exchange of arguments. Instead, we require that
\begin{align}
    &D(p_\theta || p_{\theta'}) \geq 0 \, ,\nonumber \\
    &D(p_\theta || p_{\theta'}) = 0 \iff p_\theta = p_{\theta'} \, .
    \label{eq:divergence_prop12}
\end{align}
If $D$ is differentiable in parameters then we can take $\theta' = \theta + \delta \theta$ and expand in $\delta \theta$:
\begin{align}
D(p_\theta || p_{\theta + \delta \theta}) = \frac{1}{2} g_{ij} \delta \theta_i \delta \theta_j + \mathcal{O}(\delta \theta^3) \, ,
\end{align}
where $g$ is positive-semidefinite due to \eqref{eq:divergence_prop12}. On the other hand, if $g_{ij}$ has a zero eigenvalue, the associated flat direction would correspond to infinitesimal changes in the UV theory that do not change the IR observables. Irrelevant deformations and unconstrained parameter combinations give such effects. If we take $\theta$ to correspond only to relevant or marginal deformations, we ensure that $g_{ij}$ is positive definite, locally defining a Riemannian metric. The properties \eqref{eq:divergence_prop12} together with positive-definite $g$ make $D$ a \emph{divergence}, information theoretically.

In common parlance, if a small change in UV parameters induces a small (large) change in IR observables, we say that change is natural (unnatural). In our setting we see this clearly in a diagonal basis of parameters $\delta \varphi_i$
\begin{equation}
D(p_\varphi|| p_{\varphi + \delta \varphi}) \simeq \frac12 \sum_i \lambda^{(g)}_i (\delta \varphi_i)^2 + \mathcal O(\delta\varphi^3)
\end{equation}
with the implication that directions in UV parameter space with larger eigenvalues $\lambda^{(g)}_i$ are less natural.

\bigskip 

We must introduce a final notion before arriving at our naturalness measure. A statistic $T(x)$ is called \emph{sufficient} for $\theta$ if, after compressing the data $x$ to $T(x)$, one has not lost any information about $\theta$. 
Concretely, sufficiency means that the likelihood admits a factorisation
\begin{equation}
p_\theta(x)=g_\theta\!\big(T(x)\big)\,h(x),
\end{equation}
where $h$ is independent of $\theta$. Equivalently, the conditional distribution of the microscopic data given the statistic,
\begin{equation}
p_\theta\!\big(x\,\big|\,T(x)\big),
\end{equation}
is $\theta$-independent. All $\theta$-dependence flows through $T(x)$. Sufficient statistics can be very relevant in physics, when a subset of the observables captures all of the information about the model parameters. At the most basic level, the mean and standard deviation of a Gaussian are sufficient statistics for all its moments, and hence for the full distribution. In free field theory, the one- and two-point functions are sufficient to determine all higher-point functions via Wick's theorem, since they fully characterise the underlying Gaussian measure. In conformal field theory, the spins, scaling dimensions, and OPE coefficients are sufficient to determine all correlators via the conformal bootstrap, since the crossing symmetry equations uniquely fix the spectrum and couplings of the theory~\cite{Rattazzi:2008pe}. More generally, the natural parameters of any exponential family are sufficient statistics for the full distribution, a class that includes many of the distributions encountered in statistical mechanics.

As sufficient statistics naturally arise and capture information about IR observables, we require a fine-tuning measure that is insensitive to different choices of sufficient IR observables. We do not restrict to cases where the observables are sufficient statistics, but require that the measure behaves correctly when they are.

\subsection{The Fisher Information Matrix $I_{ij}$}

Under these assumptions, Chentsov's theorem~\cite{chentsov1982statistical, amari2016information, cover2006elements} fixes $g$: the only metric invariant under sufficient statistics (up to rescaling) is the Fisher Information Metric (FIM) $g_{ij} = I_{ij}$. This provides a fundamental motivation for considering Fisher information, or something closely related to it, as a naturalness measure. 
We will now derive this metric from a divergence.

There are several divergences that lead to the metric referenced in Chentsov's theorem. The canonical choice is the Kullback-Leibler (KL) divergence,
\begin{equation}
    D_{\mathrm{KL}}(P\|Q) = \int d^n x\, P(x) \log\!\left(\frac{P(x)}{Q(x)}\right),
\end{equation}
which can be interpreted as the ``surprise'' when events are distributed according to $Q(x)$ but one had anticipated $P(x)$. While intuitive, the KL divergence is not symmetric in its arguments, so $D_{\mathrm{KL}}(P\|Q) \neq D_{\mathrm{KL}}(Q\|P)$ in general. For this reason we instead work with the Jensen--Shannon (JS) divergence,
\begin{equation}
    D_{\mathrm{JS}}(P\|Q) = \frac{1}{2}D_{\mathrm{KL}}(P\|Q) + \frac{1}{2}D_{\mathrm{KL}}(Q\|P),
\end{equation}
which is the natural symmetrisation of the KL divergence and therefore treats the two distributions on an equal footing. Unlike the KL divergence, $\sqrt{D_{\mathrm{JS}}}$ satisfies the triangle inequality making it more natural to interpret as a distance between distributions.

We want to consider how sensitive our proposed distribution on observables is to small changes in the model parameters, as illustrated in figure~\ref{fig:distributions}. To this end, we compute the JS divergence between $\rho_\theta(x)$ and $\rho_{\theta+\delta\theta}(x)$, and expand in $\delta\theta^i$,
\begin{equation}
    D_{\mathrm{JS}}(\rho_\theta \| \rho_{\theta+\delta\theta}) = \frac{1}{2} I_{ij}(\theta)\, \delta\theta^i \delta\theta^j + \mathcal{O}(\delta\theta^3),
\end{equation}
where the zeroth- and first-order terms vanish identically, and the FIM $I_{ij}$, is given by
\begin{equation}
    I_{ij}(\theta) = \int d^n x\, \rho_\theta(x) \left(\frac{\partial \log \rho_\theta(x)}{\partial \theta^i}\right) \left(\frac{\partial \log \rho_\theta(x)}{\partial \theta^j}\right).
\end{equation}
The FIM is a measure of how much information the distribution on observables $\rho_\theta(x)$ carries about the model parameters $\theta$. More generally, it may be obtained from the quadratic expansion of any $f$-divergence.

For an experimental distribution, the FIM elements $I_{ij}$ are large whenever the observations are highly informative about the parameters $\theta^i$ and $\theta^j$, which can arise for several distinct reasons:
\begin{enumerate}
    \item the observables are measured to high precision;
    \item the observables are highly sensitive to the input parameters, i.e.\ $\partial X^a / \partial \theta^i \gg X^a$; or
    \item the model has large cancellations amongst parameters -- for instance, $X^a \sim \theta^i$ and $X^b \sim \theta^j - \theta^i$, yet $\langle X^b \rangle \ll \langle X^a \rangle$.
\end{enumerate}
The first case is clearly not a form of fine-tuning, which is precisely why we proposed to use a theoretical distribution rather than an experimental one in the introduction. For a theoretical distribution this dependence shows up as dependence of $I_{ij}$ on the regulator which must be removed, as we discuss below. The second case also requires some care, as the condition $\partial X^a / \partial \theta^i \gg X^a$ depends on the choice of parametrisation, and a reparametrisation of the model can render large derivatives small or vice versa. In essence, one must decide a priori on a set of dimensionless parameters, usually guided by EFT arguments, for which a prior centred on $\mathcal O(1)$ is considered most natural~\cite{Wetterich:1983bi}. Such an assumption is unavoidable in any discussion of naturalness. We shall return to this point in section~\ref{subsec:QCD}, where we discuss dimensional transmutation as a concrete example.

With these observations in hand, we now turn to constructing a fine-tuning measure based on the FIM, subject to the requirement that the final result be independent of the regulator $\sigma$ introduced in the previous section.

\subsection{The Fine-Tuning Matrix $\mathcal F_{ij}$}
In this section we define a matrix $\mathcal{F}_{ij}$, related to the Fisher matrix $I_{ij}$, from which the regulator $\sigma$ has been removed. To do so, we note that if $\rho_\theta(x)$ depends on the observables $x^a$ only through the combination
\begin{equation}
    y^a = \frac{x^a - X^a(\theta)}{\sigma},
\end{equation}
we can define $P_\theta(y) = \sigma^n\rho_\theta(x)$. For distributions of this form, a short calculation shows that the Fisher matrix can be written as
\begin{align}
    I_{ij}(\theta) &= \int d^n y \, \frac{1}{P_\theta}
    \sum_{a,b} \frac{1}{\sigma^2} \frac{\partial P_\theta}{\partial y^a} \frac{\partial X^a}{\partial \theta^i} \frac{\partial P_\theta}{\partial y^b} \frac{\partial X^b}{\partial \theta^j} 
    \,,
\end{align}
This makes it clear that the Fisher matrix elements scale as $I_{ij} \sim \sigma^{-2}$. We now wish to take $\sigma \to 0$, and so we must remove the dependence on the regulator $\sigma$. This is achieved simply by defining a new fine-tuning matrix without the factors of $\sigma^{-1}$,
\begin{align}
    \mathcal{F}_{ij}(\theta) = \int d^n y \, \frac{1}{P_\theta} \sum_{a,b} \frac{\partial P_\theta}{\partial y^a} \frac{\partial X^a}{\partial \theta^i} \frac{\partial P_\theta}{\partial y^b} \frac{\partial X^b}{\partial \theta^j} \, .
    \label{eq:FT_mat}
\end{align}
$\mathcal{F}_{ij}$ retains only the sensitivity of the predictions $X^a(\theta)$ to the model parameters $\theta^i$ and has a smooth limit as $\sigma \to 0$.\footnote{For a Gaussian distribution $\mathcal{F}$ is independent of $\sigma$.} Furthermore, since $\mathcal{F}_{ij}$ differs from $I_{ij}$ only by an overall rescaling, any intuition regarding the FIM carries over directly to $\mathcal{F}_{ij}$.

If we assume a Gaussian distribution then one can further show:
\begin{align}
    I_{ij}(\theta) &= \int d^n y\, \frac{P_\theta}{\sigma^2} \sum_{a,b} y^a y^b \frac{\partial X^a}{\partial \theta^i}  \frac{\partial X^b}{\partial \theta^j}, \nonumber   \\
    &= \frac{1}{\sigma^2}\sum_{a,b}\delta^{ab} \frac{\partial X^a}{\partial \theta^i} \frac{\partial X^b}{\partial \theta^j},\\
    &= \frac{1}{\sigma^2}\sum_a \frac{\partial X^a}{\partial \theta^i} \frac{\partial X^a}{\partial \theta^j} \, . \nonumber   
\end{align}
As mentioned in the introduction, we focus on Gaussians for simplicity, but one is free to make other choices. Then the fine-tuning matrix becomes:
\begin{align}
    \mathcal{F}_{ij}(\theta) = \sigma^2I_{ij} =  \sum_{a} \frac{\partial X^a}{\partial \theta^i} \frac{\partial X^a}{\partial \theta^j} \, .
    \label{eq:FT_mat}
\end{align}

\begin{figure}[t]
    \centering
    \includegraphics[width=0.5\textwidth]{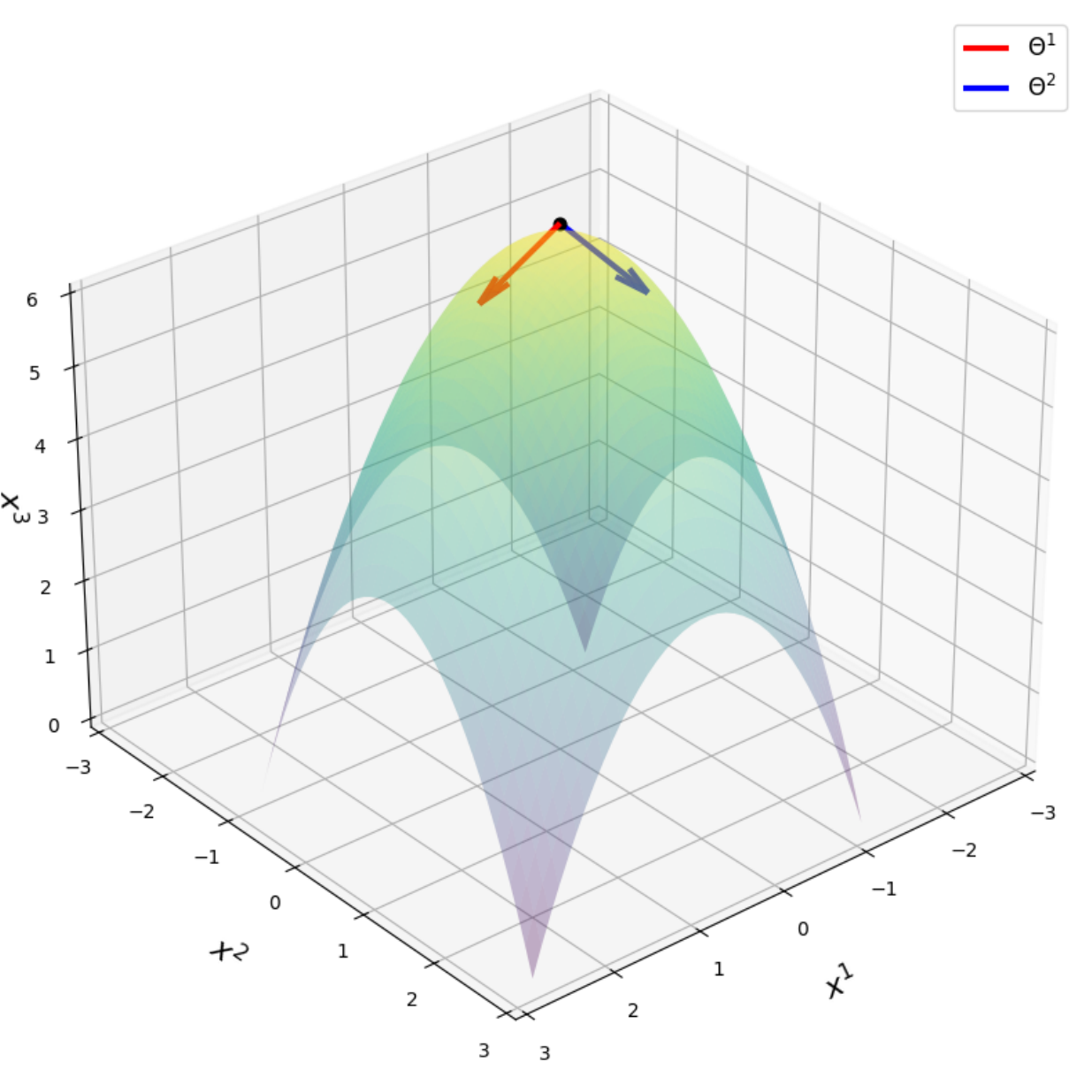}
    \caption{The submanifold of admissible observations $X(\theta)$ embedded in the three-dimensional observable space $\{x^1, x^2, x^3\}$ for a two-parameter model. The parameters $\theta^1$ and $\theta^2$ are coordinates on this surface, and the fine-tuning matrix $\mathcal{F}_{ij}$ encodes its geometry as the pullback of the Euclidean metric from the ambient observable space.}
    \label{fig:embedding}
\end{figure}

The matrix $\mathcal{F}_{ij}$ differs from the Fisher matrix $I_{ij}$ only in the removal of the $\sigma^{-2}$ factors. In the case that $n \geq N$ and we use a Gaussian distribution, one can identify $\mathcal{F}_{ij}$ as the pullback of the flat metric on observable space along the map $\theta \mapsto X(\theta)$, which provides a natural geometric interpretation of this matrix. The image of this map defines an $N$-dimensional submanifold of observable space, representing the locus of observations admissible by the model, and $\mathcal{F}_{ij}$ encodes the geometry of this submanifold as seen from parameter space. This situation is seen in figure~\ref{fig:embedding}. This raises a natural question: why choose the Euclidean metric on observable space? While this is ultimately a choice, it is difficult to motivate any metric that is not maximally symmetric, since a metric with varying curvature would implicitly privilege certain regions of observable space over others. This introduces precisely the kind of hidden assumption we are trying to avoid. The flat metric is the simplest maximally symmetric choice, though we will mention another choice in the next subsection. 

Regardless of whether we are in the context where this geometric interpretation holds, the eigenvalues of $\mathcal{F}_{ij}$ measure how sensitive the predictions $X^a(\theta)$ are to changes in the model parameters, independently of the precision to which the observables are measured. A large eigenvalue indicates a direction in parameter space along which the observables vary rapidly, while a vanishing eigenvalue indicates a direction to which the observables are entirely insensitive. It is precisely this stretching of the map $\theta \mapsto X(\theta)$ that we identify with fine-tuning, and we shall therefore use the non-zero eigenvalues of $\mathcal{F}_{ij}$ as our measure.

\subsection{Relation to Barbieri-Giudice Measure}
In the special case of a single observable $x$ and a single model parameter $\theta$, the fine-tuning matrix $\mathcal{F}_{ij}$ reduces to the scalar
\begin{align}
    \mathcal{F}(\theta) = \left(\frac{\partial X}{\partial \theta}\right)^2.
\end{align}
If one takes logarithms of both the observable and the parameter, this is precisely the fine-tuning measure proposed by Barbieri \& Giudice in ref.~\cite{Barbieri:1987fn},
\begin{align}
    c(\theta) = \left(\frac{\partial \log X(\theta)}{\partial \log \theta}\right)^2.
\end{align}
The fine-tuning measure~\eqref{eq:FT_mat} is not only more general, but the off-diagonal elements encode correlations between parameters that a single scalar measure cannot capture.

When our geometric interpretation holds we note that using the logarithmic observables corresponds to an alternative metric on observables. It corresponds to equipping observable space with the metric $\sum_a (dx^a/x^a)^2$, which is also maximally symmetric but treats observables on a logarithmic scale.

\section{Illustrative Examples}\label{sec:example}

In this section we present several illustrative examples and demonstrate that our fine-tuning matrix $\mathcal{F}_{ij}$ in each case reproduces the behaviour expected from intuitive notions of naturalness.

The first example is dimensional transmutation, which shows that large hierarchies can be generated without fine-tuning. This example also highlights that whether a theory appears fine-tuned depends on the choice of parametrisation of the UV theory. This choice can be viewed as implicitly selecting a prior on the UV parameter space, an assumption that underlies all discussions of naturalness.

We then consider the Wilson--Fisher fixed point in the $O(N)$ model, which possesses a relevant mass parameter $m$ and an irrelevant coupling $\lambda$. Reaching the fixed point requires tuning the mass, while the coupling flows there automatically. This is reflected in $\mathcal{F}_{ij}$, where one eigenvalue diverges as one flows into the IR, while the other vanishes.

Next, we analyse a two-scalar model in which a heavy scalar is integrated out at the UV scale $v$, leaving an IR theory containing a single light scalar of mass $m_h \ll v$. The fine-tuning associated with the light scalar mass diverges as $(v/m_h)^2$, providing a simple realisation of the hierarchy problem.

Finally, we consider the electron Yukawa coupling and show how technical naturalness protects a small parameter from fine-tuning, manifesting in our framework as an eigenvalue of $\mathcal{F}_{ij}$ that remains of order unity despite the large range of scales involved.

\subsection{Dimensional Transmutation/QCD}

\label{subsec:QCD}

A key consideration of any measure of fine-tuning is that it should treat observables that are highly \textit{sensitive} to the model parameters differently to fine-tuned parameters~\cite{Anderson:1994dz}. An example of this common in solutions to fine-tuning problems is that of dimensional transmutation, where the ratio of a UV to IR scale depends exponentially on the UV parameters, such as in QCD. We consider the example of QCD in this section, and show that whether a model is considered fine-tuned depends on the choice of parameters. We argue that this amounts to a choice of prior on the UV parameter space. Such a choice is inevitable in any discussion of fine-tuning, as we are always comparing the allowed region of parameter space with a predefined notion of what is a `natural' region of parameters.

The QCD scale, $\Lambda$, depends on the UV scale, $\mu_0$, and the strong coupling, $\alpha_3 (\mu_0)$, at that scale:
\begin{align}
    \Lambda = \mu_0 \exp \left( -\frac{1}{b_3 \alpha_3 (\mu_0)} \right) \, ,
\end{align}
where $b_3$ is the $\beta$-function coefficient, which can be seen in the running of $\alpha(\mu)$
\begin{equation}
    \frac{1}{\alpha(\mu)} = \frac{1}{\alpha(\mu_0)} - b_3\log\left(\frac{\mu_0}{\mu}\right),
\end{equation}
where, for $N_c$ colours and $N_f$ flavours, we have
\begin{equation}
    b_3 = \frac{1}{2\pi} \left( \frac{11}{3}N_c -\frac{2}{3}N_f  \right) \, .
\end{equation}

In order to discuss dimensionless quantities, we take $X = \log (\Lambda/\mu_0)$ to be our observable, and assume that the distribution on $X$ is a Gaussian. We can then calculate the fine-tuning in $\alpha_3 (\mu_0)$ using the measure~\eqref{eq:FT_mat}
\begin{align}
    \mathcal{F} (\alpha_3 (\mu_0)) 
    &= \frac{1}{b_3^2 \alpha_3^4 (\mu_0)}  \, .
\end{align}
This becomes large for small $b_3 \alpha_3^2$, which is not the behaviour we want. We also note that if we had taken $\log(\alpha_3)$ to be the fundamental parameter, so that $\mathcal{F}$ reproduces the Barbieri-Giudice measure, we would find
\begin{align}
    c (\alpha_3 (\mu_0)) = \frac{1}{b_3^2 \alpha_3^2 (\mu_0)} \, .
\end{align}
So $c$ also becomes large as $\alpha_3 (\mu_0) \to 0$, although the divergence on $\alpha_3(\mu_0)$ is milder.

Ref.~\cite{Anderson:1994dz} pointed out this issue with the Barbieri-Giudice measure. The authors proposed a new fine-tuning measure $\gamma(\theta)$, derived from $c(\theta)$ by dividing by the average value of $c$:
\begin{align}
    \gamma (\theta) =c(\theta) \left( \frac{\int_\Omega d\theta \, c(\theta)}{\int_\Omega d\theta} \right)^{-1}\, .
\end{align}
This method requires assuming an additional choice of the allowed region of parameter $\Omega$ space to integrate over, equivalent to introducing an explicit assumption about what is a natural allowed range of parameter values. 

Instead of introducing an averaging procedure, the unwanted apparent fine-tuning can be removed by changing variables to $\varphi = 1/(b_3\alpha_3)$. As $1/\alpha_3$ is the coefficient of the gauge kinetic term, one can argue this is a more fundamental variable than $\alpha_3$ itself. Furthermore, $\varphi$ is precisely the variable that appears naturally upon integrating the one-loop beta function, since the solution $1/\alpha_3(\mu) = 1/\alpha_3(\mu_0) - b_3\log(\mu_0/\mu)$ shows that $\varphi$ runs linearly with $\log\mu$, which is reflected in the simple form of the beta function:
\begin{align}
    \frac{\partial \varphi}{\partial \log \mu} = 1 \,.
\end{align}
If we take the fine-tuning in the parameter $\varphi$, we find that
\begin{align}
    \mathcal{F} (\varphi) 
    &= 1 \, .
\end{align}

This example highlights an important feature of any fine-tuning measure: it is not invariant under reparametrisation, since the degree of fine-tuning in a parameter necessarily depends on how that parameter is defined. This is not a weakness of our approach but an unavoidable feature of any discussion of naturalness, which always implicitly assumes a preferred set of UV variables. One can think of this as a choice of prior centred on $\mathcal{O}(1)$ values in the UV parameter space. In the dimensional transmutation example, parametrising the model with $\alpha_3$ assumes that $\mathcal{O}(1)$ values of $\alpha_3$ are natural, while parametrising with $\varphi = 1/(b_3\alpha_3)$ assumes that, at scale $\mu_0$, $\mathcal{O}(1)$ values of $\varphi$ are natural. The question of whether a theory is fine-tuned is therefore inseparable from the question of which variables one considers natural in the UV, a point that applies equally to the Barbieri--Giudice measure and all of its descendants.

\subsection{Wilson-Fisher}

In this section we consider the $O(N)$ model in $d = 4 -\epsilon$ dimensions. The theory has two parameters, a mass $m$ and a quartic coupling $\lambda$, and we assume a distribution $p_\theta$ centred around the Wilson-Fisher fixed point $m=0, \, \lambda = \lambda_* = 8\pi^2\epsilon/(N+8)$. The RG evolution of both operators is
\begin{align}
    \kappa (\mu) &\equiv \frac{m (\mu)^2 }{\mu_0^2} = \kappa_0 \left(\frac{\mu_0}{\mu} \right)^{y_m} \nonumber
    \\
     \delta \lambda (\mu) &\equiv \lambda (\mu) - \lambda_\star =\left( \lambda_0 - \lambda_\star \right)\left(\frac{\mu_0}{\mu} \right)^{y_\lambda}
\end{align}
where $\kappa$ is a dimensionless parameter equal to the mass in units of the UV scale $\mu_0$, and $\kappa_0, \, \lambda_0$ are the values of the couplings at $\mu = \mu_0$. For the $O(N)$ model in $d=4-\epsilon$ the quantities that determine the evolution under RG are
\begin{align}
    &\lambda_\star = \frac{8\pi^2 \epsilon }{N+8} \nonumber
    \\
    &y_m = 1-\frac{(N+2)}{N+8} \frac{\epsilon}{2} + \mathcal{O}(\epsilon^2)
    \\
    &y_\lambda = - \epsilon + \mathcal{O}(\epsilon^2) \nonumber
\end{align}
so $y_m >0$ indicates that the mass is a relevant parameter which needs to be tuned to reach the fixed point, while $\lambda - \lambda_\star$ is irrelevant as $y_\lambda<0$ so naturally flows to the fixed point in the IR. We thus expect that the fine-tuning in $\kappa $ will become large as the hierarchy $\mu_0 /\mu $ increases, while the fine-tuning in $\delta \lambda$ will go to zero.

As discussed above, we assume the distribution to be a Gaussian, 
\begin{align}
    \rho_{\kappa_0, \lambda_0}(x, y)=  
    \frac{1}{2\pi \sigma^2} \exp \left(- \frac{\left(x - \sqrt{\kappa (\mu)} \right)^2}{2 \sigma^2} - \frac{\left(y - \delta \lambda (\mu)\right)^2}{2 \sigma^2} \right)  \, ,
\end{align}
where $\kappa$ appears inside a square root because we measure the mass directly, not $m^2$, although this does not affect our results. The fine-tuning matrix for this model is diagonal, with eigenvalues\footnote{We note that if we had used $\kappa$ as our observable instead of $\sqrt{\kappa}$, we would have $\mathcal F_{\kappa \kappa} = \left( \mu_0 /\mu \right)^{2y_m} $, and our conclusions still apply.}
\begin{align}
    &\mathcal{F}_{\kappa \kappa} = \frac{1}{4\kappa_0}\left( \frac{\mu_0}{\mu} \right)^{y_m}  \, ,
    &&\mathcal{F}_{\lambda \lambda} = \left( \frac{\mu_0}{\mu} \right)^{2y_\lambda}  \, .
\end{align}
As we move further into the IR, $\mu \ll \mu_0$, $\mathcal{F}_{\lambda \lambda}$ vanishes, while $\mathcal{F}_{\kappa \kappa}$ diverges for a given $\kappa_0$, indicating that the mass parameter in the $O(N)$ model must be fine-tuned in order to reach the Wilson-Fisher fixed point.

\subsection{Heavy Scalar Model}

In this section we consider a model with two scalars, a light $h$ and a heavy $\Phi$, with potential:
\begin{align}
    V_{\rm uv}(h, \Phi) = \frac{m^2}{2} h^2 + \frac{\lambda_1}{4} h^4 +  \frac{\lambda_2}{4} \left( \Phi^2 - v^2 \right)^2 - \frac{\kappa}{2} h^2 \Phi^2 \, .
\end{align}
If $v \gg m_h$, where $m_h$ is the mass of $h$ we can integrate out $\Phi$ at tree level by setting it equal to its expectation value, $\Phi^2 = v^2 + \frac{ \kappa}{\lambda_2 } h^2 $. After defining the dimensionless $\gamma = m^2/v^2$, we get a potential in the IR for $h$:
\begin{align}
    V_{\rm ir} (h) &= \frac{(\gamma - \kappa) v^2}{2} h^2 + \frac{(\lambda_1 - \kappa/ \lambda_2) }{4} h^4 \, , \nonumber
    \\
    &= \frac{m_h^2 }{2} h^2 + \frac{ \lambda }{4} h^4 \, .
\end{align}
This model is fine-tuned if $m_h \ll v$ as this would require a significant cancellation between $m^2$ and $\kappa v^2$, so that $\gamma - \kappa \ll 1$.

Working at tree-level and again assuming a Gaussian, we have the probability distribution
\begin{align}
    \rho_{\gamma, \kappa, \lambda_1, \lambda_2}(x, y) = \frac{1}{2\pi \sigma^2} 
    \exp  \left( - \frac{(x - m_h/v)^2}{2 \sigma^2} - \frac{\left(y - \lambda \right)^2}{2 \sigma^2} \right)
\end{align}
From this we can calculate the fine-tuning matrix to be 
\begin{align}
\renewcommand{\arraystretch}{1.7}
    \mathcal{F} = 
    \left( 
    \begin{matrix}
        \frac{1}{4 \epsilon} & -\frac{1}{4 \epsilon} & 0 & 0  \\
        -\frac{1}{4 \epsilon} & \frac{1}{4 \epsilon} + 1/\lambda_2^{2} & -1/\lambda_2  & -\kappa/\lambda_2^3   \\
        0 & -1/\lambda_2  & 1 &  \kappa/\lambda_2^2 \\
        0 & - \kappa/\lambda_2^3  & \kappa/\lambda_2^2 &  \kappa^2/\lambda_2^4 \\
    \end{matrix}\right) \, ,
\end{align}
where we have defined $\epsilon \equiv \gamma - \kappa$. There are two vanishing eigenvalues of $\mathcal{F}$, as we only measure two quantities but the underlying model has four parameters. We are interested in the case where $h$ is light, and therefore $\epsilon\ll1$. As such, expanding in $\epsilon$, the two nonzero eigenvalues are
\begin{align}
    &1 + \frac{1}{2\lambda_2^2} + \frac{\kappa^2 }{\lambda_2^4} + \mathcal O(\epsilon)\, ,
    &&\frac{1}{2\epsilon} + \frac{1}{2\lambda_2^2 }  + \mathcal O(\epsilon) \, .
\end{align}
For $\kappa, \lambda_2 \sim \mathcal{O}(1)$ the first eigenvalue is $\mathcal{O}(1)$, reflecting the fact that there is no fine-tuning in the quartic coupling. The other observable is the mass, which is fine-tuned when there is a large hierarchy of scales, $\epsilon \ll 1$, as evidenced by the second eigenvalue becoming large in this limit.

\subsection{Electron Mass \& Technical Naturalness}

Finally, we show how technical naturalness protects a small parameter from fine-tuning by considering the example of the electron Yukawa, $y_e$. Technical naturalness states that a small parameter should be considered natural if a symmetry is restored when it is set to zero; for Yukawa couplings this symmetry is the chiral rotation of fermions~\cite{tHooft:1979rat}. The consequence is that the RG evolution of a technically natural parameter is proportional to the parameter itself, as reflected in the $\beta$-function for the electron Yukawa,
\begin{align}
    \frac{dy_e}{d\log(\mu/\mu_0)} = \frac{y_e}{16\pi^2} \left[3\sum_q y_q^2 + \sum_l y_l^2 + \frac{3y_e^2}{2}- \frac{9g_2^2}{4} - \frac{15g_1^2}{4}\right],
\end{align}
where $g_1$ and $g_2$ are the hypercharge and $SU(2)_L$ gauge couplings; the sums over $q$ and $l$ refer to summing over quark and lepton flavours, respectively.

To simplify the solution for $y_e$, we approximate the couplings in the square brackets as constant in $\mu$, giving
\begin{align}
    \frac{dy_e}{d\log(\mu/\mu_0)} &= \frac{c_e\, y_e}{16\pi^2}, \nonumber \\
    c_e &\simeq 3y_t^2 - \frac{3}{4}\left(3g_2^2 + 5g_1^2\right) \approx 1.6,
\end{align}
using $y_t = 1$, $g_2 = 0.63$, $g_1 = 0.34$. The solution is then
\begin{align}
    y_e(\mu) = y_e(\mu_0) \left(\frac{\mu_0}{\mu}\right)^{\frac{c_e}{16\pi^2}},
\end{align}
from which the fine-tuning matrix evaluates to
\begin{align}
    \mathcal{F}(y_e) = \left(\frac{\mu_0}{\mu}\right)^{\frac{c_e}{16\pi^2}}.
\end{align}
For sufficiently small $\mu$, $y_e$ appears fine-tuned, as it is a relevant coupling, analogous to the mass parameter in the Wilson--Fisher example. However, taking $\mu_0 = 10^{19}\,\text{GeV}$, $\mu = 91\,\text{GeV}$, and $c_e \approx 1.6$ gives
\begin{align}
    \mathcal{F}(y_e) \approx 1.48,
\end{align}
which is order unity, confirming that the electron Yukawa is not fine-tuned despite the large hierarchy of scales involved, because the exponent is a sufficiently small parameter.

\section{Conclusions}\label{sec:conc}

In this work we have developed an information-theoretic method of quantifying naturalness in physical theories. We constructed a fine-tuning matrix $\mathcal{F}$ in eq.~\eqref{eq:FT_mat}, whose eigenvalues determine how fine-tuned parameters of the UV theory are, given a probability distribution $\rho_\theta$ on observables. Compared to Bayesian measures of fine-tuning, using $\mathcal{F}$ gives information about the fine-tuning in each parameter of the UV theory, rather than being a global measure. It also allows us to reinterpret the Barbieri-Giudice measure as the pullback of a metric on observable space along the map $\log\theta \mapsto \log X(\theta)$.

We showed that using the eigenvalues of $\mathcal{F}$ as a measure of fine-tuning gives intuitive results for a variety of different models. With our measure, models where IR scales are generated by dimensional transmutation, technically natural parameters and couplings which flow to a stable fixed point value are all deemed to not be fine-tuned and therefore natural. By contrast, parameters which flow to an unstable fixed point and light scalars coupled to heavy fields are shown to be fine-tuned. While these results must be reproduced by any reasonable measure of fine-tuning, these examples highlight the robustness of our approach. A more detailed analysis using our measure to compare levels of tuning in more complete BSM theories such as the MSSM is left to future work. 

\section*{Acknowledgements}
TRH and JH are supported by the National Science Foundation under Cooperative Agreement PHY-2019786 (The NSF AI Institute for Artificial Intelligence and Fundamental
Interactions, http://iaifi.org/). JH is supported by NSF Award PHY-2209903. MN is supported by NSF Award PHY-2310717 and would like to acknowledge GRASP Initiative funding provided by Harvard University. 
TRH and MN would like to thank Prateek Agrawal for formulating the hierarchy problem to them in a form that inspired this work.

\appendix

\bibliographystyle{utphys}
\bibliography{main.bib}{}

@article{Susskind:1978ms,
    author = "Susskind, Leonard",
    title = "{Dynamics of Spontaneous Symmetry Breaking in the Weinberg-Salam Theory}",
    reportNumber = "SLAC-PUB-2142",
    doi = "10.1103/PhysRevD.20.2619",
    journal = "Phys. Rev. D",
    volume = "20",
    pages = "2619--2625",
    year = "1979"
}

@article{Dimopoulos:1981zb,
    author = "Dimopoulos, Savas and Georgi, Howard",
    title = "{Softly Broken Supersymmetry and SU(5)}",
    reportNumber = "HUTP-81/A022",
    doi = "10.1016/0550-3213(81)90522-8",
    journal = "Nucl. Phys. B",
    volume = "193",
    pages = "150--162",
    year = "1981"
}

@article{Fowlie:2024nhs,
    author = "Fowlie, Andrew and Herrera, Gonzalo",
    title = "{Precise interpretations of traditional fine-tuning measures}",
    eprint = "2406.03533",
    archivePrefix = "arXiv",
    primaryClass = "hep-ph",
    doi = "10.1103/PhysRevD.111.015020",
    journal = "Phys. Rev. D",
    volume = "111",
    number = "1",
    pages = "015020",
    year = "2025"
}

@article{Wilson:1971dc,
    author = "Wilson, Kenneth G. and Fisher, Michael E.",
    title = "{Critical exponents in 3.99 dimensions}",
    doi = "10.1103/PhysRevLett.28.240",
    journal = "Phys. Rev. Lett.",
    volume = "28",
    pages = "240--243",
    year = "1972"
}

@article{tHooft:1979rat,
    author = "'t Hooft, Gerard",
    editor = "'t Hooft, Gerard and Itzykson, C. and Jaffe, A. and Lehmann, H. and Mitter, P. K. and Singer, I. M. and Stora, R.",
    title = "{Naturalness, chiral symmetry, and spontaneous chiral symmetry breaking}",
    reportNumber = "PRINT-80-0083 (UTRECHT)",
    doi = "10.1007/978-1-4684-7571-5_9",
    journal = "NATO Sci. Ser. B",
    volume = "59",
    pages = "135--157",
    year = "1980"
}

@article{Nilles:1983ge,
    author = "Nilles, Hans Peter",
    title = "{Supersymmetry, Supergravity and Particle Physics}",
    reportNumber = "UGVA-DPT-1983-12-412",
    doi = "10.1016/0370-1573(84)90008-5",
    journal = "Phys. Rept.",
    volume = "110",
    pages = "1--162",
    year = "1984"
}

@article{Haber:1984rc,
    author = "Haber, Howard E. and Kane, Gordon L.",
    title = "{The Search for Supersymmetry: Probing Physics Beyond the Standard Model}",
    reportNumber = "UM-HE-TH-83-17, SCIPP-85-47",
    doi = "10.1016/0370-1573(85)90051-1",
    journal = "Phys. Rept.",
    volume = "117",
    pages = "75--263",
    year = "1985"
}

@article{Dimopoulos:1979es,
    author = "Dimopoulos, Savas and Susskind, Leonard",
    editor = "Zichichi, A.",
    title = "{Mass Without Scalars}",
    reportNumber = "CU-TP-147, ITP-626-STANFORD",
    doi = "10.1016/0550-3213(79)90364-X",
    journal = "Nucl. Phys. B",
    volume = "155",
    pages = "237--252",
    year = "1979"
}

@article{Kaplan:1983sm,
    author = "Kaplan, David B. and Georgi, Howard and Dimopoulos, Savas",
    title = "{Composite Higgs Scalars}",
    reportNumber = "HUTP-83/A079",
    doi = "10.1016/0370-2693(84)91178-X",
    journal = "Phys. Lett. B",
    volume = "136",
    pages = "187--190",
    year = "1984"
}

@article{Weinberg:1988cp,
    author = "Weinberg, Steven",
    editor = "Hsu, Jong-Ping and Fine, D.",
    title = "{The Cosmological Constant Problem}",
    reportNumber = "UTTG-12-88",
    doi = "10.1103/RevModPhys.61.1",
    journal = "Rev. Mod. Phys.",
    volume = "61",
    pages = "1--23",
    year = "1989"
}

@article{Susskind:2003kw,
    author = "Susskind, Leonard",
    editor = "Carr, Bernard J.",
    title = "{The Anthropic landscape of string theory}",
    eprint = "hep-th/0302219",
    archivePrefix = "arXiv",
    pages = "247--266",
    month = "2",
    year = "2003"
}

@article{Arkani-Hamed:2004ymt,
    author = "Arkani-Hamed, Nima and Dimopoulos, Savas",
    title = "{Supersymmetric unification without low energy supersymmetry and signatures for fine-tuning at the LHC}",
    eprint = "hep-th/0405159",
    archivePrefix = "arXiv",
    doi = "10.1088/1126-6708/2005/06/073",
    journal = "JHEP",
    volume = "06",
    pages = "073",
    year = "2005"
}

@article{Arkani-Hamed:1998jmv,
    author = "Arkani-Hamed, Nima and Dimopoulos, Savas and Dvali, G. R.",
    title = "{The Hierarchy problem and new dimensions at a millimeter}",
    eprint = "hep-ph/9803315",
    archivePrefix = "arXiv",
    reportNumber = "SLAC-PUB-7769, SU-ITP-98-13",
    doi = "10.1016/S0370-2693(98)00466-3",
    journal = "Phys. Lett. B",
    volume = "429",
    pages = "263--272",
    year = "1998"
}

@article{Randall:1999ee,
    author = "Randall, Lisa and Sundrum, Raman",
    title = "{A Large mass hierarchy from a small extra dimension}",
    eprint = "hep-ph/9905221",
    archivePrefix = "arXiv",
    reportNumber = "MIT-CTP-2860, PUPT-1860, BUHEP-99-9",
    doi = "10.1103/PhysRevLett.83.3370",
    journal = "Phys. Rev. Lett.",
    volume = "83",
    pages = "3370--3373",
    year = "1999"
}

@article{Ellis:1986yg,
    author = "Ellis, John R. and Enqvist, K. and Nanopoulos, Dimitri V. and Zwirner, F.",
    title = "{Observables in Low-Energy Superstring Models}",
    reportNumber = "CERN-TH-4350-86",
    doi = "10.1142/S0217732386000105",
    journal = "Mod. Phys. Lett. A",
    volume = "1",
    pages = "57",
    year = "1986"
}

@article{Barbieri:1987fn,
    author = "Barbieri, Riccardo and Giudice, G. F.",
    title = "{Upper Bounds on Supersymmetric Particle Masses}",
    reportNumber = "CERN-TH-4825/87",
    doi = "10.1016/0550-3213(88)90171-X",
    journal = "Nucl. Phys. B",
    volume = "306",
    pages = "63--76",
    year = "1988"
}

@article{Allanach:2006jc,
    author = "Allanach, B. C.",
    title = "{Naturalness priors and fits to the constrained minimal supersymmetric standard model}",
    eprint = "hep-ph/0601089",
    archivePrefix = "arXiv",
    reportNumber = "DAMTP-2006-5",
    doi = "10.1016/j.physletb.2006.02.052",
    journal = "Phys. Lett. B",
    volume = "635",
    pages = "123--130",
    year = "2006"
}

@article{Cabrera:2008tj,
    author = "Cabrera, M. E. and Casas, J. A. and Ruiz de Austri, R.",
    title = "{Bayesian approach and Naturalness in MSSM analyses for the LHC}",
    eprint = "0812.0536",
    archivePrefix = "arXiv",
    primaryClass = "hep-ph",
    reportNumber = "IFT-UAM-CSIC-08-81",
    doi = "10.1088/1126-6708/2009/03/075",
    journal = "JHEP",
    volume = "03",
    pages = "075",
    year = "2009"
}

@article{Anderson:1994dz,
    author = "Anderson, Greg W. and Castano, Diego J.",
    title = "{Measures of fine tuning}",
    eprint = "hep-ph/9409419",
    archivePrefix = "arXiv",
    reportNumber = "MIT-CTP-2350",
    doi = "10.1016/0370-2693(95)00051-L",
    journal = "Phys. Lett. B",
    volume = "347",
    pages = "300--308",
    year = "1995"
}

@article{Anderson:1994tr,
    author = "Anderson, Greg W. and Castano, Diego J.",
    title = "{Naturalness and superpartner masses or when to give up on weak scale supersymmetry}",
    eprint = "hep-ph/9412322",
    archivePrefix = "arXiv",
    reportNumber = "MIT-CTP-2369",
    doi = "10.1103/PhysRevD.52.1693",
    journal = "Phys. Rev. D",
    volume = "52",
    pages = "1693--1700",
    year = "1995"
}

@article{Athron:2007ry,
    author = "Athron, Peter and Miller, D. J.",
    title = "{A New Measure of Fine Tuning}",
    eprint = "0705.2241",
    archivePrefix = "arXiv",
    primaryClass = "hep-ph",
    doi = "10.1103/PhysRevD.76.075010",
    journal = "Phys. Rev. D",
    volume = "76",
    pages = "075010",
    year = "2007"
}

@article{Fichet:2012sn,
    author = "Fichet, Sylvain",
    title = "{Quantified naturalness from Bayesian statistics}",
    eprint = "1204.4940",
    archivePrefix = "arXiv",
    primaryClass = "hep-ph",
    doi = "10.1103/PhysRevD.86.125029",
    journal = "Phys. Rev. D",
    volume = "86",
    pages = "125029",
    year = "2012"
}

@article{Fowlie:2014xha,
    author = "Fowlie, Andrew",
    title = "{CMSSM, naturalness and the ''fine-tuning price'' of the Very Large Hadron Collider}",
    eprint = "1403.3407",
    archivePrefix = "arXiv",
    primaryClass = "hep-ph",
    doi = "10.1103/PhysRevD.90.015010",
    journal = "Phys. Rev. D",
    volume = "90",
    pages = "015010",
    year = "2014"
}

@article{Kim:2013uxa,
    author = "Kim, Doyoun and Athron, Peter and Bal{\'a}zs, Csaba and Farmer, Benjamin and Hutchison, Elliot",
    title = "{Bayesian naturalness of the CMSSM and CNMSSM}",
    eprint = "1312.4150",
    archivePrefix = "arXiv",
    primaryClass = "hep-ph",
    doi = "10.1103/PhysRevD.90.055008",
    journal = "Phys. Rev. D",
    volume = "90",
    number = "5",
    pages = "055008",
    year = "2014"
}

@article{Clarke:2016jzm,
    author = "Clarke, Jackson D. and Cox, Peter",
    title = "{Naturalness made easy: two-loop naturalness bounds on minimal SM extensions}",
    eprint = "1607.07446",
    archivePrefix = "arXiv",
    primaryClass = "hep-ph",
    doi = "10.1007/JHEP02(2017)129",
    journal = "JHEP",
    volume = "02",
    pages = "129",
    year = "2017"
}

@article{Wetterich:1983bi,
    author = "Wetterich, C.",
    title = "{Fine Tuning Problem and the Renormalization Group}",
    reportNumber = "CERN-TH-3528",
    doi = "10.1016/0370-2693(84)90923-7",
    journal = "Phys. Lett. B",
    volume = "140",
    pages = "215--222",
    year = "1984"
}

@article{Maity:2015rfa,
    author = "Maity, Reevu and Mahapatra, Subhash and Sarkar, Tapobrata",
    title = "{Information Geometry and the Renormalization Group}",
    eprint = "1503.03978",
    archivePrefix = "arXiv",
    primaryClass = "cond-mat.stat-mech",
    doi = "10.1103/PhysRevE.92.052101",
    journal = "Phys. Rev. E",
    volume = "92",
    number = "5",
    pages = "052101",
    year = "2015"
}

@inproceedings{Gaite:1996vk,
    author = "Gaite, Jose C.",
    title = "{Relative entropy in field theory, the H theorem and the renormalization group}",
    booktitle = "{3rd International Conference on Renormalization Group (RG 96)}",
    eprint = "hep-th/9610040",
    archivePrefix = "arXiv",
    month = "8",
    year = "1996"
}

@article{Casini:2016udt,
    author = "Casini, Horacio and Teste, Eduardo and Torroba, Gonzalo",
    title = "{Relative entropy and the RG flow}",
    eprint = "1611.00016",
    archivePrefix = "arXiv",
    primaryClass = "hep-th",
    doi = "10.1007/JHEP03(2017)089",
    journal = "JHEP",
    volume = "03",
    pages = "089",
    year = "2017"
}

@article{Cotler:2022fze,
    author = "Cotler, Jordan and Rezchikov, Semon",
    title = "{Renormalization group flow as optimal transport}",
    eprint = "2202.11737",
    archivePrefix = "arXiv",
    primaryClass = "hep-th",
    doi = "10.1103/PhysRevD.108.025003",
    journal = "Phys. Rev. D",
    volume = "108",
    number = "2",
    pages = "025003",
    year = "2023"
}

@article{Berman:2022uov,
    author = "Berman, David S. and Klinger, Marc S.",
    title = "{The Inverse of Exact Renormalization Group Flows as Statistical Inference}",
    eprint = "2212.11379",
    archivePrefix = "arXiv",
    primaryClass = "hep-th",
    reportNumber = "QMUL-PH-22-40",
    doi = "10.3390/e26050389",
    journal = "Entropy",
    volume = "26",
    number = "5",
    pages = "389",
    year = "2024"
}

@article{Stout:2021ubb,
    author = "Stout, John",
    title = "{Infinite Distance Limits and Information Theory}",
    eprint = "2106.11313",
    archivePrefix = "arXiv",
    primaryClass = "hep-th",
    month = "6",
    year = "2021"
}

@article{Stout:2022phm,
    author = "Stout, John",
    title = "{Infinite Distances and Factorization}",
    eprint = "2208.08444",
    archivePrefix = "arXiv",
    primaryClass = "hep-th",
    month = "8",
    year = "2022"
}

@book{amari2016information,
  title={Information geometry and its applications},
  author={Amari, Shun-ichi},
  year={2016},
  publisher={Springer}
}

@book{cover2006elements,
  title={Elements of information theory (wiley series in telecommunications and signal processing)},
  author={Cover, Thomas M and Thomas, Joy A},
  year={2006},
  publisher={Wiley-interscience}
}

@article{Rattazzi:2008pe,
    author = "Rattazzi, Riccardo and Rychkov, Vyacheslav S. and Tonni, Erik and Vichi, Alessandro",
    title = "{Bounding scalar operator dimensions in 4D CFT}",
    eprint = "0807.0004",
    archivePrefix = "arXiv",
    primaryClass = "hep-th",
    doi = "10.1088/1126-6708/2008/12/031",
    journal = "JHEP",
    volume = "12",
    pages = "031",
    year = "2008"
}

@book{chentsov1982statistical,
  title={Statistical decision rules and optimal inference},
  author={Chentsov, Nikolai Nikolaevich},
  year={1982},
  publisher={American Mathematical Society}
}

@article{Balasubramanian:2014bfa,
    author = "Balasubramanian, Vijay and Heckman, Jonathan J. and Maloney, Alexander",
    title = "{Relative Entropy and Proximity of Quantum Field Theories}",
    eprint = "1410.6809",
    archivePrefix = "arXiv",
    primaryClass = "hep-th",
    doi = "10.1007/JHEP05(2015)104",
    journal = "JHEP",
    volume = "05",
    pages = "104",
    year = "2015"
}

@article{Fowler:2021oje,
    author = "Fowler, Ram{\'o}n and Heckman, Jonathan J.",
    title = "{Misanthropic entropy and renormalization as a communication channel}",
    eprint = "2108.02772",
    archivePrefix = "arXiv",
    primaryClass = "hep-th",
    doi = "10.1142/S0217751X22501093",
    journal = "Int. J. Mod. Phys. A",
    volume = "37",
    number = "16",
    pages = "2250109",
    year = "2022"
}

\end{document}